%% file: article.tex
\documentclass[sigconf]{acmart}

\usepackage{booktabs} 

\usepackage{graphicx}
\usepackage{color}
\usepackage{soul}
\usepackage{todonotes}
\usepackage{hyperref}
\usepackage{listings}
\usepackage{amsmath}

\definecolor{mygreen}{rgb}{0,0.6,0}
\definecolor{mygray}{rgb}{0.5,0.5,0.5}
\definecolor{mymauve}{rgb}{0.58,0,0.82}

\setcopyright{rightsretained}

\usepackage{array}
\newcolumntype{L}[1]{>{\raggedright\let\newline\\\arraybackslash\hspace{0pt}}m{#1}}
\newcolumntype{C}[1]{>{\centering\let\newline\\\arraybackslash\hspace{0pt}}m{#1}}
\newcolumntype{R}[1]{>{\raggedleft\let\newline\\\arraybackslash\hspace{0pt}}m{#1}}

\newif\iffinal

\iffinal
  \newcommand\yadu[1]{}
  \newcommand\kyle[1]{}
  \newcommand\eamon[1]{}
\else
  \newcommand\yadu[1]{{\color{blue}[Yadu: #1]}}
  \newcommand\kyle[1]{{\color{red}[Kyle: #1]}}
  \newcommand\eamon[1]{{\color{purple}[Eamon: #1]}}
\fi

\acmDOI{10.475/123_4}

\acmISBN{123-4567-24-567/08/06}

\acmConference[ScienceCloud'17]{Workshop on Scientific Cloud Computing}{June 2017}{Washington, DC USA} 
\acmYear{2017}
\copyrightyear{2017}

\acmPrice{15.00}

\begin{document}
\title{Enabling Interactive Analytics of Secure Data using Cloud Kotta}

\author{Yadu N. Babuji, Kyle Chard, and Eamon Duede}
\affiliation{%
  \institution{Computation Institute, University of Chicago and Argonne National Laboratory}
  \streetaddress{5735 S Ellis Ave}
  \city{Chicago} 
  \state{Illinois} 
  \postcode{60637}
}
\email{{yadu, chard, eduede}@uchicago.edu}

\renewcommand{\shortauthors}{Y. Babuji et al.}

\begin{abstract}
Research, especially in the social sciences and humanities, is increasingly
reliant on the application of data science methods to analyze large amounts
of (often private) data. Secure data enclaves provide a solution for managing and analyzing private data. However, such enclaves do not readily support
discovery science---a form of exploratory or interactive analysis by which researchers 
execute a range of (sometimes large) analyses in an iterative and collaborative
manner. 
The batch computing model offered by many data enclaves is well suited to executing
large compute tasks; however it is far from ideal for day-to-day discovery science.
As researchers must submit jobs to queues and wait for results, the high latencies inherent in queue-based, batch computing systems hinder interactive
analysis.
In this paper we describe how we have augmented the Cloud Kotta secure data enclave
to support collaborative and interactive analysis of sensitive data.
Our model uses Jupyter notebooks as a flexible analysis environment 
and Python language constructs to support the execution of arbitrary 
functions on private data within this secure framework.
\end{abstract}

\newcommand{\NAMENS} {\textsc{Cloud Kotta}} 
\newcommand{\NAME} {\textsc{Cloud Kotta }}

\maketitle

\input{introduction}
\input{background}
\input{arch}

\input{applications}
\input{related}
\input{summary}


\section*{Acknowledgments}

The authors would like to thank Nathan Bartley and Alexander Belikov for testing the system in its early stages of development.

\bibliographystyle{ACM-Reference-Format}
\bibliography{references}

\end{document}

%% file: introduction.tex
\section{Introduction}

Regardless of domain, scientists are rapidly embracing data-driven science as a 
means of extracting knowledge from large amounts of data. While there are many examples
of successful research based upon big data in the biomedical~\cite{toga15big}, physical~\cite{reed15exascale}
and social~\cite{foster2015tradition} sciences, many researchers still face challenges
managing and analyzing large amounts of data. These challenges are even more complex
when the data to be analyzed is sensitive, private, or valuable. Furthermore, the increasingly common
adoption of `discovery science', an approach that centers on iterative and exploratory analysis
of large volumes of data to discover patterns and correlations, is 
beyond the reach of many researchers who lack the expertise and computational infrastructure
to support such an approach to interrogating large, sensitive data. In practice, the requirements
for analyzing sensitive, big data using discovery and data science methodologies
necessitates flexible, intuitive, and efficient infrastructure. Yet, most highly secure, big data computational infrastructure is anything but flexible and intuitive.

To address the need for the secure and scalable management and 
analysis of data, we developed the Cloud Kotta secure data enclave~\cite{babuji16secure,babuji16kotta}
Cloud Kotta provides a cloud-based, elastically scalable environment that is 
able to meet the needs of sporadic and bursty scientific analysis workloads
while removing the need for owning and operating large scale infrastructure.
It implements a fine grain access control model over managed research data 
allowing controlled access from within and outside the enclave. To address the need for reliability, scalability and collaborative access, Cloud Kotta is built upon Amazon Web Services (AWS). Over the past year, Cloud Kotta has been leveraged by dozens of researchers and students to analyze data using more than a quarter million core hours.

While Cloud Kotta has shown immense value, it, like many
other data enclaves, offers only a queue-based job submission model. 
Unfortunately, such models are not well-suited to the increasingly
common discovery science approaches used by researchers. Rather, discovery and data science approaches typically rely on lightweight scripting languages (e.g., R and Python), 
flexible data structures (e.g., dataframes), inline visualizations (to `inspect' the data), 
and exploratory statistical and machine learning algorithms.
These days, researchers rely on interactive analysis environments like Jupyter Notebooks~\cite{jupyter} 
to support iterative and collaborative analysis that marries code, equations, documentation, 
results, and visualizations. A quick and responsive environment that allows for fast, iterative
development is an ideal fit for analysis and discovery tasks. For this reason, these interactive development environments have quickly become crucial tools of
the applied data science community and are starting to gain favor across a broad range of computational sciences. However, maintaining the
high availability requirements of interactive compute resources can be expensive especially
with a large group of researches to support. Furthermore, providing interactive analysis
of secure data is particularly challenging.

In this paper, we describe a model for enabling interactive, multi-user analysis 
of secure data. We base our model on Jupyter Notebooks and Cloud Kotta to provide the security of a data enclave, scalable compute, and the interactivity required of today's discovery science. 
We have developed a Python library that enables specific functions in an analysis code, written in 
a Jupyter Notebook, to be seamlessly and securely submitted
to the Cloud Kotta execution fabric. Our approach streamlines
the interfaces between the analysis code and the execution
environment thereby offering native, synchronous Python input/output
alongside an asynchronous job-based model for long running analyses. 
To validate our model we describe several real-world use cases 
for which this system was used. 

%% file: background.tex
\section{Background and Motivation}

A gap has opened up between researchers' preferred methods for working with and interrogating data and the usability of the computational environments that host that work. Researchers are increasingly adopting highly flexible, iterative, approaches to data exploration. These approaches are most prominent in the applied data science and discovery science regimes in which large quantities of heterogeneous data are explored, smashed together, interrogated by general and specialized data mining and machine learning tools in a highly iterative process of rapid ideation and exploration. As this methodology has gained favor, researchers have attempted to leverage it in working with ever larger, more diverse, and highly sensitive data (e.g., patient, genetic, tax, commercial, etc.). At the same time, computational facilities, resources, and platforms have sought to optimize throughput, performance, and security with little regard for the interactivity or the intuitive, inventive, flights of exploratory fancy that researchers now crave. On one side of the gap, users of these facilities now have scale at the expense of flexibility and security at the expense of interactivity. On the other, users have flexibility and interactivity at the expense of scale and security.

Cloud Kotta was developed to serve the computational needs of a diverse network of scientists that rely on big data and computation to ask and answer big questions. Originally, Cloud Kotta was designed to virtually centralize the research efforts of a decentralized group around highly protected datasets. In order to carry out analysis of highly sensitive data at a distance, researchers were (initially) willing to suffer shortcomings related to interactivity and flexibility. However, iterating on models and troubleshooting bugs in the fairly rigid, queue-based job submission analysis workflows that Cloud Kotta supported proved cumbersome and, ultimately, frustrated research efforts.

For this reason, we sought to extend Cloud Kotta by developing and providing an interactive analysis environment that enables users to work directly with elastic computational resources and highly sensitive data in a flexible, secure, and interactive environment at scale. Specifically, for interactive data analysis on Cloud Kotta to be efficacious, we derived the following requirements.

The augmented system needed to: 
1) maintain high levels of security, ensuring that only permitted users
are able to access and interact with data while restricting what data can leave the system;
2) provide an intuitive interface that is both familiar to users and flexible
enough to support a broad range of analysis types;
3) seamlessly integrate with a diverse set of programming languages so that users do not
need to make significant modifications to their programs;
4) scale to support very large amounts of data and use of large compute
resources; and
5) embrace software as a service models such that users can access these
capabilities without installing software locally.  

In what follows, we sketch the Cloud Kotta architecture and then describe in detail our novel Jupyter integrations that enable interactive analysis of data within the Cloud Kotta environment. Additionally, we discuss three real-world use cases that have leveraged these capabilities to conduct science.

%% file: arch.tex
\section{Design and Architecture}

\begin{figure}
  \center
  \includegraphics[width=0.45\textwidth]{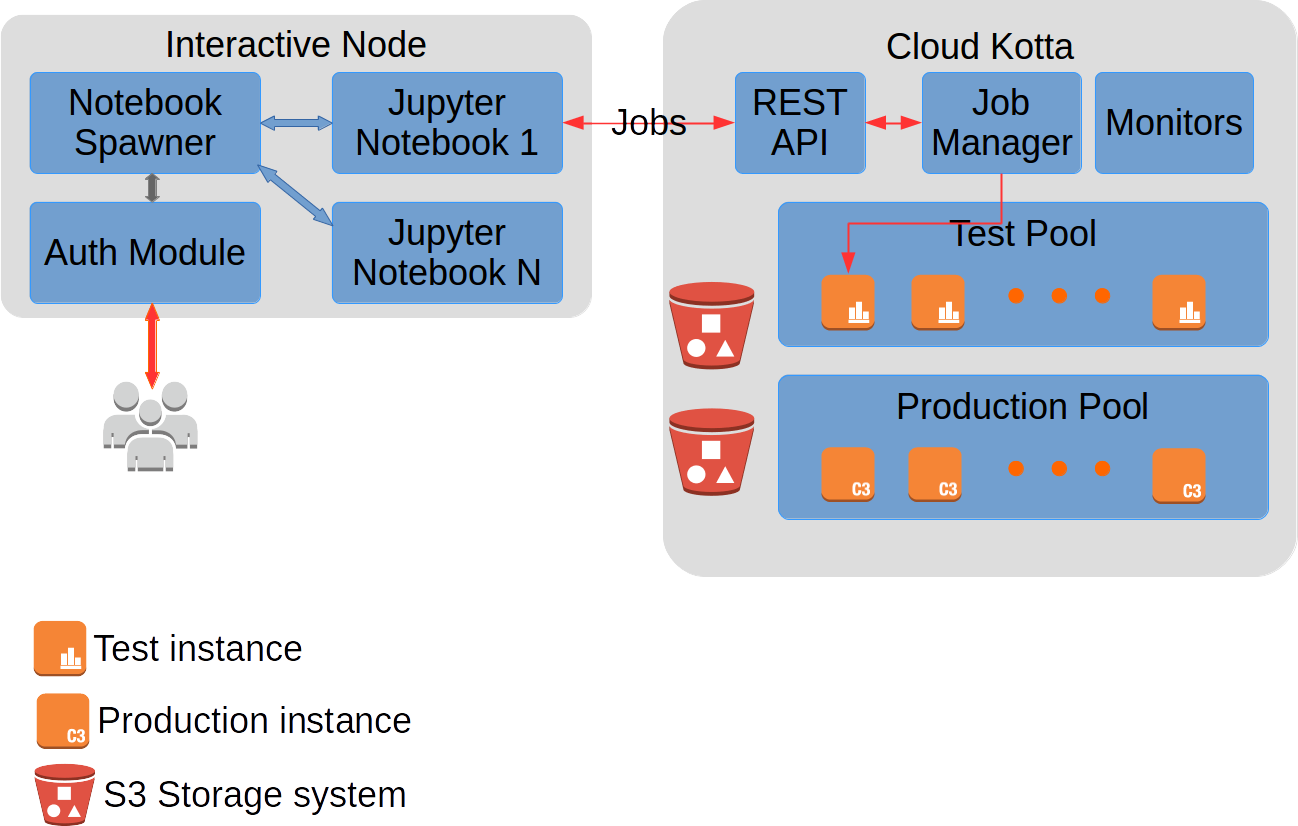}
  \caption{Architecture.}
  \label{fig:arch}
  \vspace{-1.5em}
\end{figure}

In an effort to address the requirements described in the previous section, we extended Cloud Kotta to integrate the Jupyter development environment and marry its interactive
execution model with the Cloud Kotta computation fabric. Our approach allows users to select
and analyze secure datasets while leveraging the virtually 
infinite compute and storage capacity of the cloud all within a familiar and intuitive environment. 
The architecture of our solution is depicted in \figurename~\ref{fig:arch}.

\subsection{Cloud Kotta}
Cloud Kotta is a cloud-based service for conducting collaborative research around protected datasets.
It is designed to be deployed on Amazon Web Services (AWS), leveraging highly reliable
platform services and scalable data storage and compute environments. 
Cloud Kotta leverages a suite of AWS service including 
Simple Storage Service (S3) for long term data storage, 
DynamoDB for storing fine grain job histories, 
Simple Queue Service (SQS) for rapid and reliable task distribution,
and Elastic Compute Cloud (EC2) for an elastically scalable computing environment.

Cloud Kotta stores and manages protected data in S3. This provides a
low cost model for storing large amounts of data while also providing
a fine grain access control model for individual objects. Moreover, Kotta offers
secure HTTPS interfaces for accessing data. 
When data is to be analyzed, it is moved and cached on cloud instance storage
using AWS Elastic Block Storage.
Cloud Kotta employs advanced data management policies that move data within storage 
tiers based on frequency of access. This model allows us to trade off
high availability for lower cost similar to least-recently-used caching. 
Combining these methods allows Cloud Kotta to reduce storage costs, significantly.

To satisfy bursty and potentially large-scale compute workloads
Cloud Kotta employs multiple elastic pools of compute resources. 
This allows the system to match incoming workloads so as to
provide seemingly infinite compute capacity at a fraction of the 
cost of maintaining a large persistent cluster in the cloud. 
That is, instances are only initiated when needed so that users are only charged for the resources needed for compute.
Where possible, Cloud Kotta leverages low cost \emph{spot} instances
and adopts a smart bidding system across four availability zones to ensure
that the resources provisioned are the cheapest available machines. 

To service the two broad job categories on the system, Cloud Kotta maintains a test queue that is geared
for development jobs and a separate production queue for jobs that require significant CPU and or memory resources.
The test queue is comprised of an elastic pool of \emph{t2.medium} instances, each with 2 vCPUs and 4GB of RAM. These instances
are \emph{on-demand} instances which can be provisioned in under two minutes. 
To ensure rapid response time, the test 
queue always has at least one machine on-line. The production queue, on the other hand, is usually populated with c3.8xlarge
(32vCPU, 60GB RAM) or i2.8xlarge (36vCPU, 244GB RAM) machines that are provisioned from the spot market.
These machines are slower to provision and are designed for jobs that are generally tolerant of delays.

In order to utilize these resources, Cloud Kotta provides a queue based job submission model that is accessible
through a Web GUI, REST API and a command line interface. The Web GUI is convenient for submitting single tasks
and tracking their progress, but become a hassle when dealing with a large number of jobs. The REST interface
is designed to support programmatic access for example from external applications of via scripts that aim to manage several hundreds of jobs. This
flexibility comes with the additional effort of fitting science workflows to this model. The command line
interface is useful for submitting a large number of jobs with minor variations. This is suited for bash
script based tasks. 

\subsection{Security}

Security of the hosted data is the primary concern for a secure data enclave. Cloud Kotta is architected with
this in mind, and so implements a rich, fine grain access control model. 
At the heart of the model users are assigned roles. All datasets define policies
with respect to these roles, for example defining who can access each dataset.
Policies also control what permissions individual services have. 
Every request that involves data is signed and logged such that 
access can be audited in the future.

Rather than implement a user management system, Cloud Kotta leverages 
Amazon's identity services. Cloud Kotta's web interface 
relies on Login with Amazon, an OAuth~2.0 provider for user authentication. 
This workflow allows users to authenticate with their Amazon identity. 
Cloud Kotta is given a short-term token that can be used to verify identity
and conduct operations on that user's behalf. While the authentication
tokens themselves are valid for only an hour, we create cookies to extend 
session validity to 6 hours. This approach works well for web-based sessions; 
however it is not suitable in a programmatic context. To simplify programmatic
authentication Cloud Kotta leverages refresh tokens issued through a one-time registration
process. These tokens can be refreshed by the application for as long as they
are valid. Users may revoke tokens at any point in the future. 

To support analysis of private data on elastic cloud instances Cloud
Kotta leverages a unique delegated access model. Each instance is 
created with a special role, this role provides a minimal set of 
privileges by default (e.g., to read from the queue). However, the role
allows the instance to inherent the role of the user that owns a submitted
job. In this case, the instance changes to the user's role which 
allows it to access protected data throughout the execution of the 
analysis task.  

Cloud Kotta applies best practices cloud security models to secure
the entire system. For example, S3 buckets are encrypted and 
accessible only from within a predefined virtual private cloud (VPC). 
The compute layer is isolated from the internet by a private subnet.


\subsection{Jupyter Deployment}

Jupyter is an application that allows users to author living documents
that contain code (e.g., Python, R, Julia), rich descriptive text (e.g., documentation), 
and the results of running code (e.g., text, figures). Notebooks are organized into cells,
users can execute these cells in any order and state is shared between cells. 
Jupyter notebooks are authored and executed in a web browser, using a locally
deployed web server. 
One of the reasons Jupyter has become so popular is that it provides 
a flexible and interactive model for analysis, it is intuitive to use
via a web browser, and notebooks are self-describing. It is this combination
that makes it easy to share, reproduce, and extend analyses.
However, Jupyter notebooks are limited in that they are single-user
instances, they are not designed to exploit large scale computing
infrastructure, and they do not provide support for accessing 
data securely. 

JupyterHub~\cite{jupyterhub} is a multi-user server that manages user authentication and 
can spawn multiple single-user Jupyter notebooks. We base our integration
on JupyterHub for this reason. To do so, we have deployed JupyterHub
on an EC2 instance within the Cloud Kotta system. We use JupyterHub's PAM
authentication system to authenticate users onto the login node with access
to a persistent filesystem. This authentication system is currently separate
from the system used by Cloud Kotta and therefore requires a second user
account.  We are actively working to integrate the two models. 
JupyterHub users are provided with their own (isolated) space that
can be used for creating and maintaining stateful development environments
(e.g., notebooks and temporary data). We use a notebook spawner
to create new notebook instances for users on demand. This spawner is designed
to restrict the CPU and RAM available to each notebook such that
individual users will not be able to accidentally render the service unusable.
As we will describe later, our library extensions allow users to offload analysis
tasks that require resources beyond what is allocated to an individual notebook instance. 


Since availability and reliability is critical for the Jupyter environment, it is important that the
host EC2 instance capacity matches expected workload (i.e., number of users). To address this
concern, the Cloud Kotta system administration team monitors load and can scale-up or scale-down the
instance type that is provisioned for the service.

\subsection{Kotta Interactive Library}

\begin{figure}
  \center
  \includegraphics[width=0.45\textwidth]{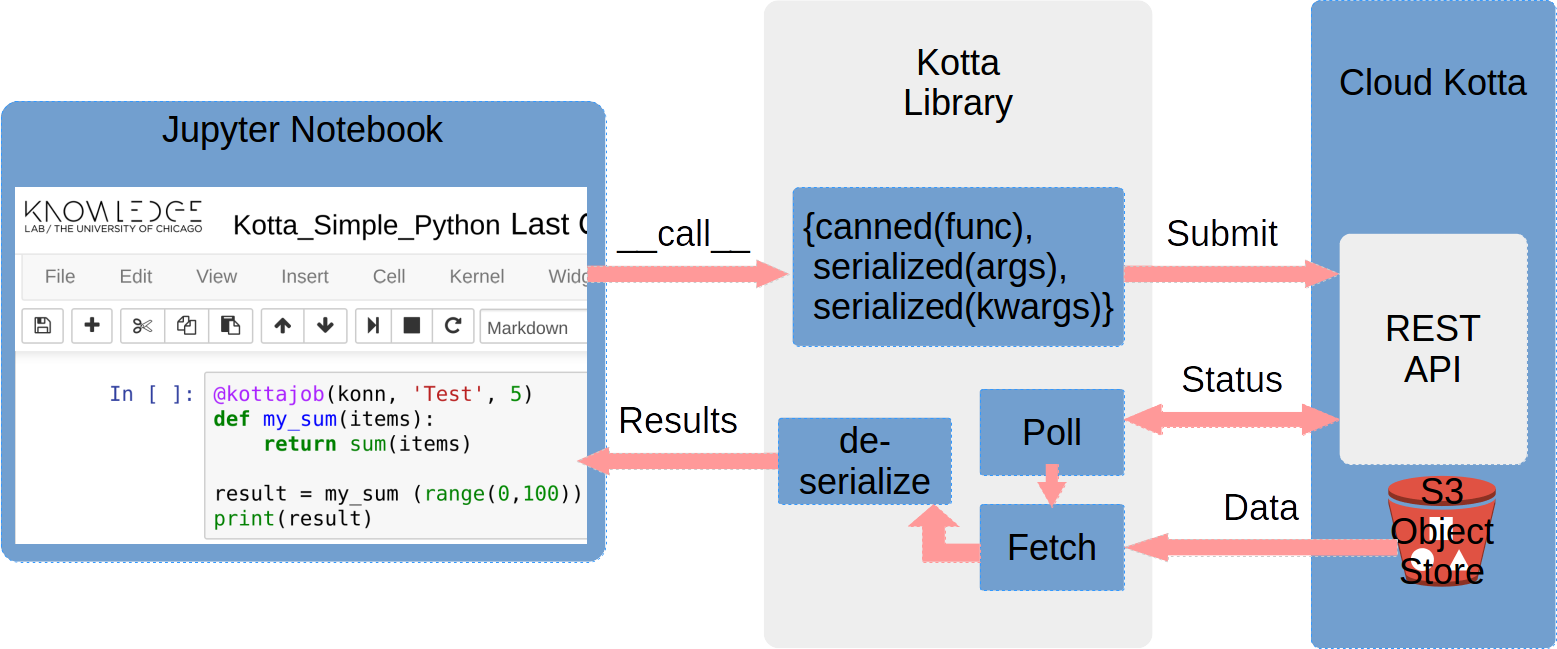}
  \caption{Kotta Library sequence diagram}
  \label{fig:kotta_lib}
  \vspace{-1.5em}
\end{figure}

The primary interface used to connect Jupyter notebooks with
Cloud Kotta is the \texttt{kotta} python library. This library enables the user to
markup python functions using a decorator. Functions decorated with the \texttt{@kottajob} decorator, when called
are executed transparently on the Cloud Kotta infrastructure. The library handles the serialization of the
arguments passed, canning of the decorated function, and deserialization of the results once they
have been computed. The decorator takes an authenticated connection object, the target Cloud Kotta queue and a walltime as
required arguments. 
The \texttt{kotta} library authenticates with Cloud Kotta using valid access tokens.
These tokens are, at present, stored in a file within the user's private file system. 
This sequence of steps taken by the \texttt{kotta} library is shown in \figurename~\ref{fig:kotta_lib}

Here is a simple example of a decorated Python function
which simply computes the sum of a list of integers:

\begin{lstlisting}
@kottajob(konn, 'Test', 5)
def my_sum(items):
   return sum(items)

result = my_sum(range(0,100))
print(result)
\end{lstlisting}

The above code snippet, when executed with the \texttt{kotta} library, 
looks to the user like any other Python script, albeit with the additional latencies from
serialization, transmission and execution on a remote compute node. The function runs in blocking mode,
therefore \texttt{my\_sum()} does not return until its results are available. The \texttt{@kottajob}
decorator can also run in non-blocking mode and that makes it simple to exploit many-task parallelism
within analysis workflows.

\begin{lstlisting}
@kottajob(konn, 'Test', 5, block=False)
def my_sqrt(items):
   import math
   return [math.sqrt(x) for x in items]

jobs = []
# Split the problem space into 5 chunks of size 20
for item in range(0,100,20):
   jobs.extend([my_sum(range(item,item+20))])

# Wait for results
print([job.result for job in jobs])
\end{lstlisting}

Since the function \texttt{my\_sqrt()} is decorated with the option \texttt{block=False}, calls to
the function return immediately with a \emph{future}. A future is a common term used to describe
an object that acts as a proxy for which a value is not yet known. 
The advantage of this approach is that users may execute several such functions, all
of which are executed in parallel.

Decoratated functions also take special keywords \texttt{inputs} and \texttt{outputs} which are
used to indicate to Cloud Kotta that special (protected) datasets must be staged-in or staged-out.
In this case, Cloud Kotta uses its secure data enclave to locate the requested dataset. Cloud Kotta
then invokes its standard security mechanisms to authorize access to the data if the user
has privileges to access the dataset. 
Since each task is handled by Cloud Kotta there's an auditable
record of every compute task that touched protected datasets.

\begin{lstlisting}
@kottajob(konn, 'Test', 10)
def file_sum(inputs=[]):
   import os
   print(os.listdir('.'))
   lines = open(os.path.basename(inputs[0]), 'r').readlines()
   data = [int(line.strip()) for line in lines]
   total = sum(data)
   length = len(data)
   return total, length

returns = file_sum(inputs=['s3://klab-jobs/1m_shuffled.txt'])
print(returns)
\end{lstlisting}

The \texttt{kotta} library by default captures both STDOUT and STDERR streams. These are accessible through
the future returned from a non-blocking call, or when a blocking call fails with an exception.
This simplifies debugging during the development process. It is also important to note that Cloud Kotta
logs the entirety of the task allowing both retrospective and prospective provenance. For tasks running
longer than 5 minutes Cloud Kotta also captures continuous CPU and Memory utilization information that can be viewed
through the Web GUI. The \texttt{@kottajob} decorator also takes a \texttt{requirements} keyword argument that
accepts a specially formatted string. This allows for installing and setting up requisite python packages
before the decorated function is executed on the compute nodes.

Finally, the library also supports specifying execution of arbitrary applications in a batch model. This allows the user to easily
plug in existing applications written in any language and to subsequently manage these applications via Python. 
This ensures that complex workflows with unusual
requirements such as ensuring that no data is fetched to the Jupyter notebook are possible.
Here is a minimal example:

\begin{lstlisting}
job = KottaJob(jobtype="script",
               jobname="Hello",
               executable='/bin/bash myscript.sh',
               script_name='myscript.sh',
               script="""/bin/bash
               echo 'Hello world'
               """)

# Submit a job with the Kotta connection object
job.submit(konn)

# Track job status
job.status(konn)

# Print stdout
print(job.stdout)
\end{lstlisting}

%% file: applications.tex
\section{Applications}

Adding interactivity to Cloud Kotta for rapid ideation and iteration was motivated by a gap between the functionality of the computational environment and the type of interrogative, discovery science adopted by researchers. This gap was particularly noticeable in work being carried out by Knowledge Lab researchers on the Clarivate Analytics (formerly the IP \& Science business of Thomson Reuters) {\it Web of Science} and the ITHAKA {\it JSTOR} datasets. This work, situated in the Computational Social Sciences and the broader Science of Science, seeks to understand the institutions and dynamics of knowledge production by modeling the scientific system from publication data~\cite{evans2011metaknowledge}. In what follows, we outline three concrete use cases that initially motivated and then benefited from our extension of Cloud Kotta.

Cloud Kotta has been used to analyze the full {\it Web of Science} raw data and to compute two measures of the scientific impact of individual journals and individual papers -- Eigenfactor and the Article Influence metrics, respectively~\cite{bergstrom2008eigenfactor}. The general analysis pipeline is as follows. The {\it Web of Science} data files are converted from their raw XML format to a structured collection of lists and dictionaries for each year. From this intermediate form, the data are processed and
aggregated in batches corresponding to moving window years (e.g., as specified by the Eigenfactor computation). 
Of course, any large computational infrastructure would have proved efficacious for this job. However, because of the sensitive nature of the data and accompanying access agreements, the use of the Cloud Kotta secure data enclave and computational infrastructure was necessitated. The ability to collocate propriatery data and compute within the same virtual private and secure cloud was also convenient, due to the size of the dataset (~100 Gb), the compute time (several hours per job), and the susceptibility of the task to parallelization (the initial dataset is split into year-labeled files). The collection of jobs was submitted en masse through the Cloud Kotta Client Python interface described in this paper and outputs were fetched in a similar manner. It is worth noting that, while the selection of machine types and OS images are hidden from the user, the system is
flexible enough that, upon request by the user, instances with, for example, more memory or an image with task specific
pre-installed libraries, can be made available --which is what occurred in this case. For future bibliometric work, users will be in a position to make minor modifications to the setup and codebase used in the Eigenfactor computation thanks to Cloud Kotta's job publication features.

In addition to parsing tens of millions of files and calculating metrics from extracted metadata, the interactive Jupyer analysis environment is ideal for generating and then iteratively interrogating networks (e.g. graphs) comprised of billions of links. Much of the work that goes into the initial analysis of networks is rather exploratory. However, when a network is as massive as, for instance, the citation network that underlies the scientific enterprise, rapid ideation is stymied by secure but rigid computational environments with queue based job submission protocols. 

In order to get a better sense of the citation network in the {\it Web of Science}, researchers needed to parse tens of millions of raw XML files, generate an edge list of all source papers and their respective citations targets, and create an adjacency list. The goal was to make it easier to directly parse, interrogate, and derive summary statistics about citation counts for any specific paper in the {\it Web of Science}. The derived, proprietary edge list citation network is 20 GB uncompressed, making it immediately a memory and disk-intensive problem for local machines or small, single nodes. Moreover, because the data is proprietary and subject to data use agreements, the network cannot be replicated and processed locally. Given that we are working with billions of citations and parallelizable algorithms for assembling the adjacency list, it makes economic and legal sense to utilize the Kotta decorator and farm the problem out to large nodes. Moreover, once the graph has been generated, basic summary statistics can be fetched directly from the Jupyter notebook by running simple network analysis functions with the Cloud Kotta decorator. In this way, even a network with billions of links can be explored in the straightforward and intuitive way that researchers have come to expect in an area of discovery science.

Finally, users have found the new, interactive analysis capabilities of Cloud Kotta to be useful in conducting research using natural language processing and text mining on big, scholarly, data. For instance, in one project, researchers seek to analyze the rhetorical framing of Adam Smith and his political philosophy from {\it The Wealth of Nations}. In order to do this, users first conducted text mining to identify terms and concepts related to Adam Smith as well as the contexts in which they appear within the body of social science literature from 1890s to the 2010s. Cloud Kotta allowed researchers to actively and iteratively filter through candidate terms and concepts and relate them to similar concepts and phrases found in Adam Smith's work. This problem required leveraging 20 GB of protected OCR'd (optical character recognition) data from JSTOR, and performing relatively expensive analyses (e.g., tf-idf, word2vec, lda, etc.) which makes the problem just out of reach for many personal, local machines, both in terms of memory and disk, as well as data use agreements.

Of course, it was possible to carry out many of the tasks required of these projects in more traditional, queue based compute environments. Several of the tasks, however, could not have been performed on more traditional data enclaves, as these have tended to be local, relatively small, and isolated machines. Nevertheless, enabling researchers to rapidly ideate and iterate on code and analysis has greatly improved the research process in each of the applications and has, as a result, sped discovery.

%% file: related.tex
\section{Related Work}

There are few systems that enable secure management and 
analysis of research data. While researchers have explored 
methods for enhancing client-side software for analysis of 
large amounts of local data~\cite{saleem14bigexcel} and 
exploited hybrid cloud models to scale analytics~\cite{abramson14hybrid},
they do not provide a flexible, integrated environment
for managing and analyzing data, and to the best of our knowledge
none support interactive analysis on protected data. Perhaps the most similar 
approach to ours is the data capsule~\cite{zheng14capsules} model used by
the Hathi Trust to support secure, non-consumptive analysis of data by leveraging 
controlled virtual machines. However, this model is designed to use virtual
machine constructs and therefore it is lower level and does not support interactive analytics via
an easy to use Jupyter notebook.

Science gateways~\cite{wikinsdiehr07gateways} have long been used to 
provide simplified, domain-specific access to large scale computing
infrastructure. However, unlike Cloud Kotta, they focus primarily
on HPC infrastructure and typically support fixed analysis types
(via a web form) and queue-based execution models. While some
gateways now leverage cloud infrastructure~\cite{madduri2014globus}
none provide the rich security policies, extensible execution model, 
or interactive analysis model provided by Cloud Kotta. 

There are several efforts to provide multi-user, interactive
analysis environments built around Jupyter notebooks.
For example, JupyterHub~\cite{jupyterhub}, the system 
we build upon here, allows multiple users to
instantiate instances of Jupyter notebooks. It is typically
deployed on a large machine and uses a proxy-based model
to forward requests to a particular Jupyter instance. 
Tmpnb~\cite{tmpnb} aims to satisfy a similar multi-user
model for running temporary notebooks. It launches Docker
containers for each notebook and proxies requests
to each container. Tmpnb has been used to provide
temporary notebooks for replicating analyses published 
in Nature~\cite{shen14notebooks}. 
Binder~\cite{binder} applies a similar model, using
Docker containers to execute Jupyter notebooks directly
from GitHub repositories.  The public deployment
is hosted on a small Google Compute Engine Cluster that can
scale with usage. 
While these systems provide on-demand interactive analysis environments,
they do so at the notebook level and focus on computational reproducibility. 
Cloud Kotta instead offers a similar environment for composing and executing notebooks
that analyze protected data. Our model is unique in that it extends these 
notebook environments via language constructs to exploit access
to secure data and use of large scale computing resources. 

Finally, there are many examples of libraries and programming 
languages that aim to simplify the use of parallel and distributed computing resources. 
In particular, IPythonParallel provides a simple model for 
enabling parallel execution of Python functions. IPythonParallel
allows user-defined decorators to be associated with functions
which are subsequently sent as parallel jobs to a predefined 
execution system. This model provides no support for data management, 
implementing security models at the granularity of functions, or autoscaling in a cloud
environment. It would require considerable effort to integrate such frameworks
with the Cloud Kotta security fabric. 

%
%
%

%% file: summary.tex
\section{Summary}

We have described the enhancements we have made to the Cloud Kotta
secure data enclave to support interactive data analytics on protected data. 
The motivation for our work was based on the needs of an increasingly common
class of researcher: those who utilize exploratory data science to analyze large, protected datasets.
To address the needs of these researchers we described how we have integrated Jupyter notebooks 
with Cloud Kotta to fulfill the significant gap between interactive and queue-based systems. 
Our approach relies on JupyterHub to enable multi-user Jupyter
environments and the creation of a lightweight Python library that supports semi-transparent execution
of code functions using a Python decorator. 
Initial experiences with this platform have been positive,  
several researchers have now adopted this system in their every-day 
research. 

Our future work is primarily focused on completing the integration
of our authentication and authorization systems. In so doing, we will
simplify user experience by enabling them to use the same identities
in both environments and by transparently enabling connections from 
Jupyter notebooks to Cloud Kotta. As a secondary goal we will extend
the \texttt{kotta} library to include support for dependency management.
This support will allow users of our system to compose more complex
workflows comprised of independent steps. 

%

%% file: article.bbl

\begin{thebibliography}{00}


\ifx \showCODEN    \undefined \def \showCODEN     #1{\unskip}     \fi
\ifx \showDOI      \undefined \def \showDOI       #1{{\tt DOI:}\penalty0{#1}\ }
  \fi
\ifx \showISBNx    \undefined \def \showISBNx     #1{\unskip}     \fi
\ifx \showISBNxiii \undefined \def \showISBNxiii  #1{\unskip}     \fi
\ifx \showISSN     \undefined \def \showISSN      #1{\unskip}     \fi
\ifx \showLCCN     \undefined \def \showLCCN      #1{\unskip}     \fi
\ifx \shownote     \undefined \def \shownote      #1{#1}          \fi
\ifx \showarticletitle \undefined \def \showarticletitle #1{#1}   \fi
\ifx \showURL      \undefined \def \showURL       #1{#1}          \fi
\providecommand\bibfield[2]{#2}
\providecommand\bibinfo[2]{#2}
\providecommand\natexlab[1]{#1}
\providecommand\showeprint[2][]{arXiv:#2}

\bibitem[\protect\citeauthoryear{??}{bin}{2017}]%
        {binder}
 \bibinfo{year}{2017}\natexlab{}.
\newblock \bibinfo{title}{Binder}.  (\bibinfo{year}{2017}).
\newblock
\showURL{%
\url{http://mybinder.org//[lastaccessed,March2017}}


\bibitem[\protect\citeauthoryear{??}{jup}{2017a}]%
        {jupyter}
 \bibinfo{year}{2017}\natexlab{a}.
\newblock \bibinfo{title}{Jupyter Notebook}.  (\bibinfo{year}{2017}).
\newblock
\showURL{%
\url{http://jupyter.org/[lastaccessed,March2017}}


\bibitem[\protect\citeauthoryear{??}{jup}{2017b}]%
        {jupyterhub}
 \bibinfo{year}{2017}\natexlab{b}.
\newblock \bibinfo{title}{JupyterHub}.  (\bibinfo{year}{2017}).
\newblock
\showURL{%
\url{https://github.com/jupyterhub/jupyterhub[lastaccessed,March2017}}


\bibitem[\protect\citeauthoryear{??}{tmp}{2017}]%
        {tmpnb}
 \bibinfo{year}{2017}\natexlab{}.
\newblock \bibinfo{title}{tmpnb, the temporary notebook service}.
  (\bibinfo{year}{2017}).
\newblock
\showURL{%
\url{https://github.com/jupyter/tmpnb[lastaccessed,March2017}}


\bibitem[\protect\citeauthoryear{Abramson, Horka, and Wisniewski}{Abramson
  et~al\mbox{.}}{2014}]%
        {abramson14hybrid}
\bibfield{author}{\bibinfo{person}{S. Abramson}, \bibinfo{person}{W. Horka},
  {and} \bibinfo{person}{L. Wisniewski}.} \bibinfo{year}{2014}\natexlab{}.
\newblock \showarticletitle{A Hybrid Cloud Architecture for a Social Science
  Research Computing Data Center}. In \bibinfo{booktitle}{{\em Proceedings of
  the 34th International Conference on Distributed Computing Systems Workshops
  (ICDCSW)}}. \bibinfo{pages}{45--50}.
\newblock
\showISSN{1545-0678}
\showDOI{%
\url{http://dx.doi.org/10.1109/ICDCSW.2014.32}}


\bibitem[\protect\citeauthoryear{Babuji, Chard, Gerow, and Duede}{Babuji
  et~al\mbox{.}}{2016a}]%
        {babuji16kotta}
\bibfield{author}{\bibinfo{person}{Y.~N. Babuji}, \bibinfo{person}{K. Chard},
  \bibinfo{person}{A. Gerow}, {and} \bibinfo{person}{E. Duede}.}
  \bibinfo{year}{2016}\natexlab{a}.
\newblock \showarticletitle{Cloud Kotta: Enabling secure and scalable data
  analytics in the cloud}. In \bibinfo{booktitle}{{\em IEEE International
  Conference on Big Data (Big Data)}}. \bibinfo{pages}{302--310}.
\newblock
\showDOI{%
\url{http://dx.doi.org/10.1109/BigData.2016.7840616}}


\bibitem[\protect\citeauthoryear{Babuji, Chard, Gerow, and Duede}{Babuji
  et~al\mbox{.}}{2016b}]%
        {babuji16secure}
\bibfield{author}{\bibinfo{person}{Y.~N. Babuji}, \bibinfo{person}{K. Chard},
  \bibinfo{person}{A. Gerow}, {and} \bibinfo{person}{E. Duede}.}
  \bibinfo{year}{2016}\natexlab{b}.
\newblock \showarticletitle{A secure data enclave and analytics platform for
  social scientists}. In \bibinfo{booktitle}{{\em 12th IEEE International
  Conference on e-Science (e-Science)}}. \bibinfo{pages}{337--342}.
\newblock
\showDOI{%
\url{http://dx.doi.org/10.1109/eScience.2016.7870917}}


\bibitem[\protect\citeauthoryear{Bergstrom, West, and Wiseman}{Bergstrom
  et~al\mbox{.}}{2008}]%
        {bergstrom2008eigenfactor}
\bibfield{author}{\bibinfo{person}{Carl~T Bergstrom}, \bibinfo{person}{Jevin~D
  West}, {and} \bibinfo{person}{Marc~A Wiseman}.}
  \bibinfo{year}{2008}\natexlab{}.
\newblock \showarticletitle{The Eigenfactor™ metrics}.
\newblock \bibinfo{journal}{{\em Journal of Neuroscience\/}}
  \bibinfo{volume}{28}, \bibinfo{number}{45} (\bibinfo{year}{2008}),
  \bibinfo{pages}{11433--11434}.
\newblock


\bibitem[\protect\citeauthoryear{Evans and Foster}{Evans and Foster}{2011}]%
        {evans2011metaknowledge}
\bibfield{author}{\bibinfo{person}{James~A Evans} {and}
  \bibinfo{person}{Jacob~G Foster}.} \bibinfo{year}{2011}\natexlab{}.
\newblock \showarticletitle{Metaknowledge}.
\newblock \bibinfo{journal}{{\em Science\/}} \bibinfo{volume}{331},
  \bibinfo{number}{6018} (\bibinfo{year}{2011}), \bibinfo{pages}{721--725}.
\newblock


\bibitem[\protect\citeauthoryear{Foster, Rzhetsky, and Evans}{Foster
  et~al\mbox{.}}{2015}]%
        {foster2015tradition}
\bibfield{author}{\bibinfo{person}{Jacob~G Foster}, \bibinfo{person}{Andrey
  Rzhetsky}, {and} \bibinfo{person}{James~A Evans}.}
  \bibinfo{year}{2015}\natexlab{}.
\newblock \showarticletitle{Tradition and innovation in scientists’ research
  strategies}.
\newblock \bibinfo{journal}{{\em American Sociological Review\/}}
  \bibinfo{volume}{80}, \bibinfo{number}{5} (\bibinfo{year}{2015}),
  \bibinfo{pages}{875--908}.
\newblock


\bibitem[\protect\citeauthoryear{Madduri, Chard, Chard, Lacinski, Rodriguez,
  Sulakhe, Kelly, Dave, and Foster}{Madduri et~al\mbox{.}}{2015}]%
        {madduri2014globus}
\bibfield{author}{\bibinfo{person}{Ravi Madduri}, \bibinfo{person}{Kyle Chard},
  \bibinfo{person}{Ryan Chard}, \bibinfo{person}{Lukasz Lacinski},
  \bibinfo{person}{Alex Rodriguez}, \bibinfo{person}{Dinanath Sulakhe},
  \bibinfo{person}{David Kelly}, \bibinfo{person}{Utpal Dave}, {and}
  \bibinfo{person}{Ian Foster}.} \bibinfo{year}{2015}\natexlab{}.
\newblock \showarticletitle{The {Globus Galaxies} platform: delivering science
  gateways as a service}.
\newblock \bibinfo{journal}{{\em Concurrency and Computation: Practice and
  Experience\/}} \bibinfo{volume}{27}, \bibinfo{number}{16}
  (\bibinfo{year}{2015}), \bibinfo{pages}{4344--4360}.
\newblock
\showISSN{1532-0634}
\showDOI{%
\url{http://dx.doi.org/10.1002/cpe.3486}}


\bibitem[\protect\citeauthoryear{Reed and Dongarra}{Reed and Dongarra}{2015}]%
        {reed15exascale}
\bibfield{author}{\bibinfo{person}{Daniel~A. Reed} {and} \bibinfo{person}{Jack
  Dongarra}.} \bibinfo{year}{2015}\natexlab{}.
\newblock \showarticletitle{Exascale Computing and Big Data}.
\newblock \bibinfo{journal}{{\em Commun. ACM\/}} \bibinfo{volume}{58},
  \bibinfo{number}{7} (\bibinfo{date}{June} \bibinfo{year}{2015}),
  \bibinfo{pages}{56--68}.
\newblock
\showISSN{0001-0782}
\showDOI{%
\url{http://dx.doi.org/10.1145/2699414}}


\bibitem[\protect\citeauthoryear{Saleem, Varghese, and Barker}{Saleem
  et~al\mbox{.}}{2014}]%
        {saleem14bigexcel}
\bibfield{author}{\bibinfo{person}{M.~A. Saleem}, \bibinfo{person}{B.
  Varghese}, {and} \bibinfo{person}{A. Barker}.}
  \bibinfo{year}{2014}\natexlab{}.
\newblock \showarticletitle{{BigExcel}: A web-based framework for exploring big
  data in social sciences}. In \bibinfo{booktitle}{{\em Proceedings of the IEEE
  International Conference on Big Data (Big Data)}}. \bibinfo{pages}{84--91}.
\newblock
\showDOI{%
\url{http://dx.doi.org/10.1109/BigData.2014.7004458}}


\bibitem[\protect\citeauthoryear{Shen}{Shen}{2014}]%
        {shen14notebooks}
\bibfield{author}{\bibinfo{person}{Helen Shen}.}
  \bibinfo{year}{2014}\natexlab{}.
\newblock \showarticletitle{Interactive notebooks: Sharing the code}.
\newblock \bibinfo{journal}{{\em Nature\/}} \bibinfo{volume}{515},
  \bibinfo{number}{7525} (\bibinfo{year}{2014}), \bibinfo{pages}{151--152}.
\newblock


\bibitem[\protect\citeauthoryear{Toga, Foster, Kesselman, Madduri, Chard,
  Deutsch, Price, Glusman, Heavner, Dinov, Ames, Van~Horn, Kramer, and
  Hood}{Toga et~al\mbox{.}}{2015}]%
        {toga15big}
\bibfield{author}{\bibinfo{person}{Arthur~W Toga}, \bibinfo{person}{Ian
  Foster}, \bibinfo{person}{Carl Kesselman}, \bibinfo{person}{Ravi Madduri},
  \bibinfo{person}{Kyle Chard}, \bibinfo{person}{Eric~W Deutsch},
  \bibinfo{person}{Nathan~D Price}, \bibinfo{person}{Gustavo Glusman},
  \bibinfo{person}{Benjamin~D Heavner}, \bibinfo{person}{Ivo~D Dinov},
  \bibinfo{person}{Joseph Ames}, \bibinfo{person}{John Van~Horn},
  \bibinfo{person}{Roger Kramer}, {and} \bibinfo{person}{Leroy Hood}.}
  \bibinfo{year}{2015}\natexlab{}.
\newblock \showarticletitle{Big biomedical data as the key resource for
  discovery science}.
\newblock \bibinfo{journal}{{\em Journal of the American Medical Informatics
  Association\/}} \bibinfo{volume}{22}, \bibinfo{number}{6}
  (\bibinfo{year}{2015}), \bibinfo{pages}{1126}.
\newblock
\showDOI{%
\url{http://dx.doi.org/10.1093/jamia/ocv077}}


\bibitem[\protect\citeauthoryear{Wilkins-Diehr}{Wilkins-Diehr}{2007}]%
        {wikinsdiehr07gateways}
\bibfield{author}{\bibinfo{person}{Nancy Wilkins-Diehr}.}
  \bibinfo{year}{2007}\natexlab{}.
\newblock \showarticletitle{Special Issue: Science Gateways—Common Community
  Interfaces to Grid Resources}.
\newblock \bibinfo{journal}{{\em Concurrency and Computation: Practice and
  Experience\/}} \bibinfo{volume}{19}, \bibinfo{number}{6}
  (\bibinfo{year}{2007}), \bibinfo{pages}{743--749}.
\newblock
\showISSN{1532-0634}
\showDOI{%
\url{http://dx.doi.org/10.1002/cpe.1098}}


\bibitem[\protect\citeauthoryear{Zeng, Ruan, Crowell, Prakash, and Plale}{Zeng
  et~al\mbox{.}}{2014}]%
        {zheng14capsules}
\bibfield{author}{\bibinfo{person}{Jiaan Zeng}, \bibinfo{person}{Guangchen
  Ruan}, \bibinfo{person}{Alexander Crowell}, \bibinfo{person}{Atul Prakash},
  {and} \bibinfo{person}{Beth Plale}.} \bibinfo{year}{2014}\natexlab{}.
\newblock \showarticletitle{Cloud Computing Data Capsules for
  Non-consumptiveuse of Texts}. In \bibinfo{booktitle}{{\em Proceedings of the
  5th ACM Workshop on Scientific Cloud Computing (ScienceCloud)}}.
  \bibinfo{publisher}{ACM}, \bibinfo{pages}{9--16}.
\newblock
\showISBNx{978-1-4503-2911-8}
\showDOI{%
\url{http://dx.doi.org/10.1145/2608029.2608031}}


\end{thebibliography}
